\documentclass[10pt,letterpaper,twocolumn,prl,showpacs]{revtex4-1}

\usepackage[english]{babel}
\usepackage{amsmath,amssymb}
\usepackage{graphicx}
\usepackage{epstopdf}
\usepackage{color}
\usepackage[usenames,dvipsnames]{xcolor}
\usepackage{hyperref}
\DeclareGraphicsExtensions{.eps,.pdf}

\begin{document}

\title{Mode-locked Yb-doped fiber laser emitting broadband pulses \\ at ultra-low repetition rates}

\author{Patrick Bowen$^{1,*}$, Miro Erkintalo$^{1}$, Richard Provo$^{2}$, John D. Harvey$^{1,2}$, and Neil G. R. Broderick$^{1}$}

\affiliation{[1] The Dodd Walls Centre of Photonic and Quantum Technologies, Department of Physics, University of Auckland, Auckland 1142, New Zealand
\\ [2] Southern Photonics, Level 4, 49 Symonds St, Auckland 1010, New Zealand \\ $^*$Corresponding author: pbow027@aucklanduni.ac.nz}

\begin{abstract}
We report on an environmentally stable, Yb-doped, all-normal dispersion, mode-locked fibre laser that is capable of creating broadband pulses with ultra-low repetition rates. Specifically,  through careful positioning of fibre sections in an all-PM-fibre cavity mode-locked with a nonlinear amplifying loop mirror, we achieve stable pulse trains with repetition rates as low as 506~kHz. The pulses have several nanojules of energy and are compressible down to ultrashort ($<$~500~fs) durations.
\end{abstract}

\maketitle

\noindent Mode-locked fibre lasers have numerous applications ranging from medicine to micro-machining \cite{NatureMicroscopy,NatureUltrafastFibreLaser}, which drives continued research interest in developing novel source configurations with enhanced performance in terms of pulse characteristics and reliability. In particular, a prominent area of research over the past decade has been lasers constructed entirely from all-normal dispersion (ANDi) elements \cite{ChongANDifirst,ChongANDi20nJ}. Such ANDi lasers have proven to be capable of producing orders of magnitude more pulse energy than typical soliton lasers \cite{SolitonEnergyLimit}, without the need for complex dispersion mapping schemes \cite{SolitonSimilariton}. Furthermore, the pulses they emit typically have broad spectral widths and linear chirps, allowing their temporally broad output pulses to be externally compressed to ultrashort ($<$~500~fs) durations.\looseness=-1

ANDi lasers are typically most stable with repetition rates at or above 10~MHz, i.e. for cavity lengths less than twenty metres. For longer fibre lengths (hence, lower repetition rates), instabilities arise that are dominantly driven by optical nonlinearities, namely self-phase modulation and stimulated Raman scattering. This is unfortunate, as applications such as ablative micro-machining and medical imaging, would benefit from lasers emitting sub-picosecond pulses at repetition rates considerably lower than 1~MHz \cite{MicroMachiningkHzRepRate,MicroMachiningHeat,Imaging1,Imaging2}. Achieving such low repetition rates in a fibre laser system generally requires one or more acousto-optic modulators to externally pulse pick from a source laser with a much higher repetition rate ($\sim$~10~MHz). This process is lossy by nature, expensive to implement, and adds complexity to the system. For this reason, it would be advantageous to posses a source capable of directly delivering lower repetition rates.\looseness=-1

The ANDi lasers with the lowest repetition rates are based on designs known as giant chirp oscillators (GCOs) \cite{GCOKelleherTheory}. This design was first demonstrated by Renninger et al. \cite{FirstGCO}, who used nonlinear polarisation evolution to passively mode-lock a 3~MHz laser cavity, achieving output pulses with $\sim$~5~nm spectral width that could be compressed to 670~fs in duration. More recently, a 1.7~MHz device, mode-locked using a nonlinear amplifying loop mirror (NALM), was reported by some of us \cite{ClaudeLargeChirpLaser}. The laser emitted pulses with 12~nm spectral width that were compressible to 300~fs. However, subsequent research has shown that the repetition rate of that design cannot be significantly reduced below 1.7 MHz \cite{NLRamanDriven,AntoineGCOunstable}. This is because the elongation of the cavity gives rise to enhanced stimulated Raman scattering which destabilizes the laser operation. Other GCO designs have achieved much lower repetition rates ($<$~100~kHz) by using SESAMs, carbon nanotubes, or nonlinear polarisation evolution for mode-locking \cite{GCOchirpMeasurement,64kHzRamanLaser,GCOfreespace77kHz,LinearCavityLowRepRate,GCOnonPM,LowRepRateSESAM,LowRepRateSESAMLinear}. Although these realisations do not appear to suffer from Raman-driven destabilisation, their output pulses generically possess very narrow ($<$~1 nm) spectral widths, which prevents compression to ultrashort durations and limits the range of applications these lasers can service.\looseness=-1

In this Letter, we report on a GCO that emits stable trains of linearly chirped pulses with broad ($>$~10~nm) spectral bandwidths and repetition rates as low as 506 kHz. Our design is based on the laser reported in \cite{ClaudeLargeChirpLaser}; however, we demonstrate that nonlinear instabilities can be mitigated by careful distribution of fibre segments in the cavity. To the best of our knowledge, the repetition rates achieved in our device are the lowest ever reported in any all-fibre, environmentally stable oscillator whose pulses are compressible to ultrashort durations.\looseness=-1

\begin{figure}[t]
\centering
\includegraphics[width=\linewidth]{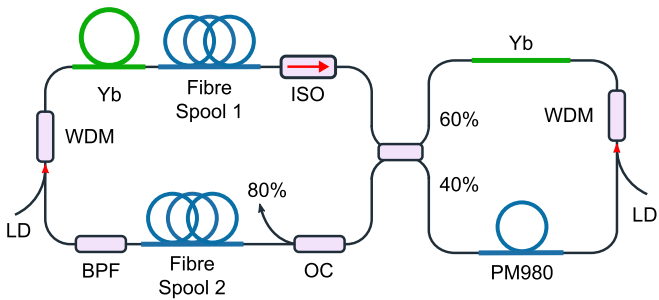}
\caption{Schematic drawing of the studied fibre laser cavity. WDM: wavelength division multiplexer, Yb: Ytterbium-doped active fibre, LD: laser diode, BPF: band-pass filter, OC: output coupler, PM980: type of passive fibre used in the NALM.}
\label{SchematicGCO}
\end{figure}

The experimental setup of the figure-eight laser under study is schematically illustrated in Fig.~\ref{SchematicGCO}. It consists of two loops: a main uni-directional section and the NALM. The main loop contains up to 374~m of fibre in two spools, a 5~m low-doped Yb gain section, an output coupler, and a narrow 1.6~nm band-pass filter centred at 1030~nm. The NALM loop is bi-directional and serves to mode-lock the laser. It contains a length of passive fibre, a short segment of highly-doped Yb fibre, and is connected to the main loop by an unequally split (60:40) central coupler. The two separately pumped gain sections (one in the main loop and the other in the NALM) provide an extra degree of freedom for mode-locking, and allow for the fine-tuning of the output pulse characteristics \cite{UoANALMreview,Claude120fsLaser,ClaudeLaser,MyPaper1}.\looseness=-1

Compared to the device reported in \cite{ClaudeLargeChirpLaser}, the key design difference that has enabled us to overcome nonlinear limitations is that, in the current design, cavity elongation is achieved by placing two fibre spools at different positions in the main cavity, as shown in Fig.~\ref{SchematicGCO}. This should be contrasted with the design in \cite{ClaudeLargeChirpLaser}, where all the SMF was placed immediately after the gain fibre in the main loop. To show explicitly the effect of splitting the SMF into two sections, we built a laser with a repetition rate of 835~kHz (cavity length $\sim$~240~m), and performed experiments in two configurations whose only difference manifested itself in SMF positioning. In the first configuration (old design, single-spool), 214 meters of SMF was placed immediately after the gain fibre in the main loop; in the second configuration (new design, double-spool), 214 meters of SMF was distributed in the two positions shown in Fig.~\ref{SchematicGCO}, with 114~m in spool 1 and 100~m in spool 2. We found that, in the first configuration, the laser could Q-switch and partially mode-lock to produce noise-like pulses \cite{AntoineGCOunstable,HorowitzNoiselikeFirst}. However, we found no combination of pump powers that would allow us to reach a stable mode-locked regime. In stark contrast, in the double-spool configuration the laser was found to easily self-start and reach stable mode-locking operation. In general the laser first emits noise like pulses when the pump lasers are switched on, but stable mode locked pulses are readily achieved by independent adjustment of the two pumps. In the case of the 240~m length laser, this results in stable mode locked operation, with an average output power of 7.7~mW, corresponding to 9.3~nJ energy per pulse.\looseness=-1





\begin{figure}[t]
\centering
\includegraphics[width=\linewidth]{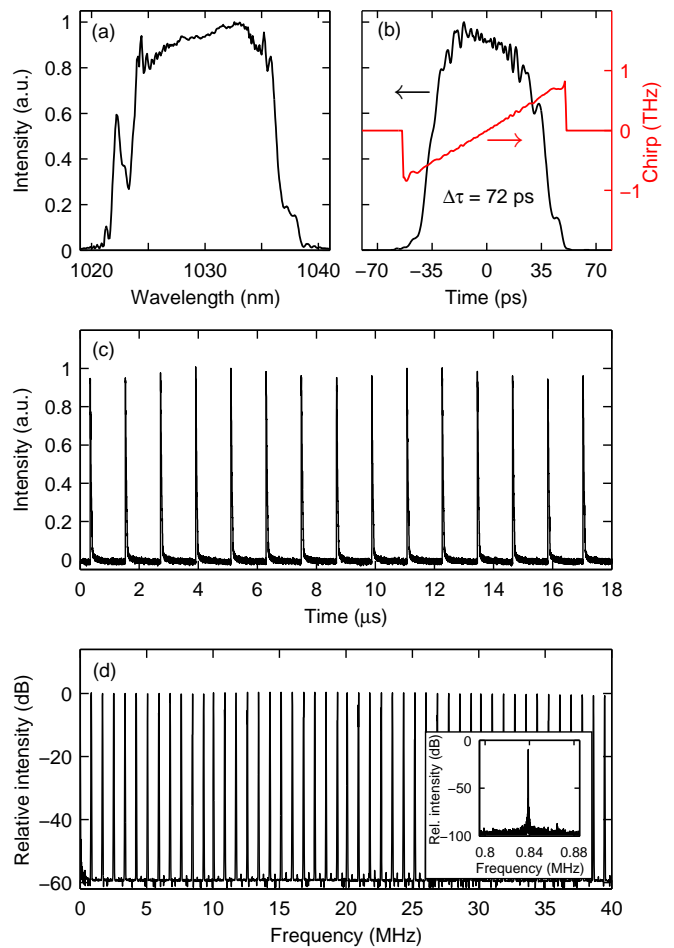}
\caption{Output characteristics of 835~kHz laser. (a) Spectrum in linear scale, (b) uncompressed pulse profile from FROG trace, (c) pulse train measured on a photodetector, and (d) corresponding RF spectrum. Inset in (d) shows a zoom around the fundamental repetition frequency.}
\label{OldGCOvsNew}
\end{figure}

Figure~\ref{OldGCOvsNew} shows spectral and temporal characteristics of the laser output in the double-spool configuration. The spectrum (measured with an optical spectrum analyser Antritzu MS9710) is shown in Fig.~\ref{OldGCOvsNew}(a), and it can be seen to be centered at 1030~nm and to have a full width at half maximum (FWHM) of about 14~nm. Moreover, the spectrum has steep edges as is characteristic for ANDi lasers \cite{ChongANDifirst,ClaudeLargeChirpLaser,AntoineGCOunstable}. Figure~\ref{OldGCOvsNew}(b) shows the temporal intensity and phase profiles of the output pulses, retrieved from a frequency-resolved optical gating (FROG) measurement performed using a commercial device (Southern Photonics HR100). The output pulses can be seen to have 72~ps duration (FWHM), and to be linearly chirped. We also note that additional autocorrelation measurements (not shown here) show no evidence of a ``coherence peak'', which further evidences that the laser is not operating in the noise-like regime. We also show in Figs.~\ref{OldGCOvsNew}(c) and \ref{OldGCOvsNew}(d) the pulse train and corresponding radio-frequency (RF) spectrum measured with a fast photodetector connected to an oscilloscope and an RF spectrum analyser, respectively. We clearly see a repetitive 835 kHz pulse train with a clean RF spectrum, as expected for stable single-pulse mode-locking.\looseness=-1


\begin{figure}[t] 
\centering
\includegraphics[width=\linewidth]{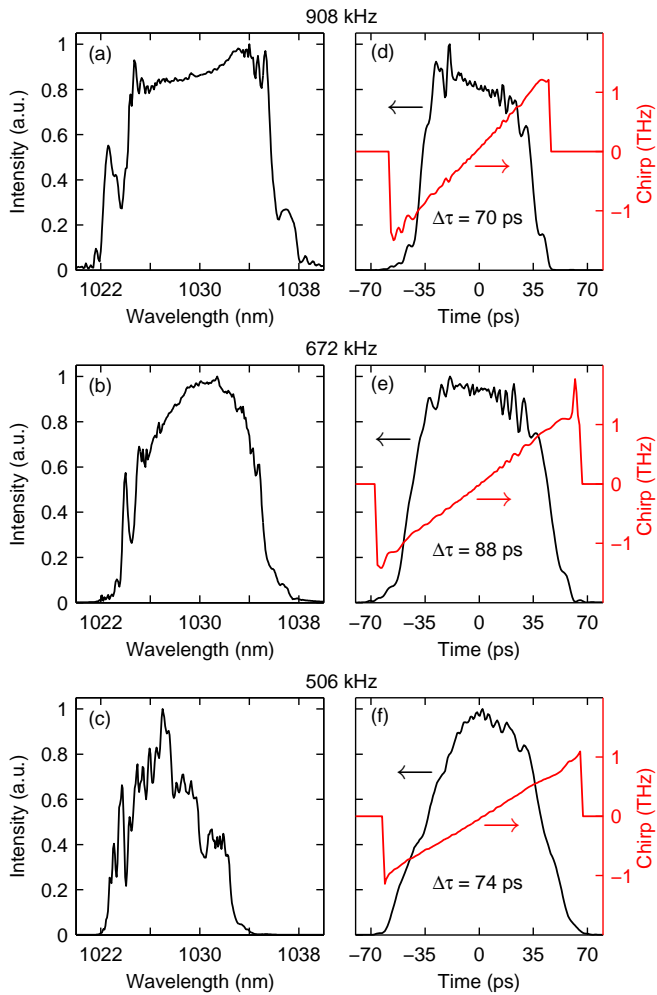}
\caption{Output pulse spectra (a--c) in linear scale, and uncompressed pulse profiles (d--f) from a FROG trace. Measurements correspond to the emitted pulses for various repetition rates of the laser cavity, shown above the figure panels. $\Delta\tau$ is the FWHM of the temporal profile.}
\label{OtherRepRates}
\end{figure}

To demonstrate the robustness of the new double-spool design, and to understand how the laser output characteristics might change as a function of the cavity length, we have tested a range of different fibre spool lengths (hence, cavity repetition rates). For each configuration, we have found that the laser can be easily mode-locked by adjusting the pump power levels, and that the output pulses always show characteristics qualitatively similar to those described above (e.g. stable trains of nanojoule pulses). Indeed, Fig.~\ref{OtherRepRates} shows measured spectral [Fig.~\ref{OtherRepRates}(a)--(c)] and temporal [Fig.~\ref{OtherRepRates}(d)--(f)] characteristics for selected cavity configurations with repetition rates as indicated (see also Table~\ref{TableSpecs}). Several conclusions can be drawn. First, for each configuration, the emitted pulses have broad spectra, with FWHM widths exceeding 10 nm. Second, the temporal profiles are also broad (and exhibit complex shapes), yet possess almost linear chirps. Third, Fig. 3(c) and (f) show that we have achieved stable mode-locking with a repetition rate as low as 506 kHz. In this configuration, the spectrum exhibits very complex fine structure, but the time-domain pulse profile nevertheless appears comparatively clean and linearly chirped. Indeed, as discussed below, the laser output can readily be externally compressed to ultrashort durations. At this point we note that the finite amount of SMF available in our laboratory has prevented us from testing configurations with even lower repetition rates.\looseness=-1

For all of the configurations that we have tested, the laser shows high environmental stability. Specifically, the laser stays mode-locked for hours without deviations in pulse characteristics, despite being operated in a weakly-controlled laboratory with large temperature variations in the range of  $\pm 3^\circ$C. Furthermore, the laser is extremely robust against mechanical stress, and its operation is not easily disturbed by e.g. human-induced vibrations or other perturbations. Of course, such environmental stability is to be expected based on the fact that the laser is constructed out of all-PM-fibre components.\looseness=-1

\begin{figure}[t]
\centering
\includegraphics[width=\linewidth]{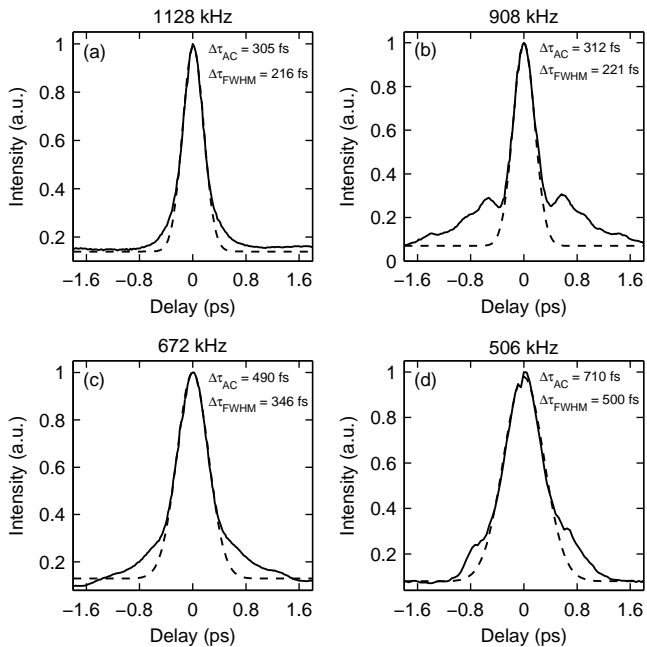}
\caption{Compressed autocorrelation traces of output pulses at various repetition rates shown above the figure panels.  $\Delta\tau_\mathrm{AC}$ is the FWHM of the autocorrelation trace, while $\Delta\tau_\mathrm{FWHM}$ is the FWHM of a corresponding Gaussian pulse.}
\label{compressedACtraces}
\end{figure}

The temporal profiles of the recovered output pulses, shown in Fig.~\ref{OldGCOvsNew}(b) and Fig.~\ref{OtherRepRates}(d--f), exhibit visibly linear chirp, suggesting they may be externally compressed close to their (femtosecond-scale) transform-limit. To directly test compressibility, we used a multi-pass transmission grating arrangement with a variable path length of 40--160~cm, and a grating line density of 1000 lines/mm. A 4-pass arrangement was used for all but the 506~kHz configuration, which required an 8-pass arrangement to avoid beam diffraction beyond the finite width of the compressor system. The compressed pulses were then characterized using an autocorrelator (APE pulseCheck) which has a better resolution than our FROG device. The results are shown in Fig.~\ref{compressedACtraces}, where we plot the autocorrelation traces corresponding to the different configurations discussed above. As can be seen, for each configuration, the compressed pulses have widths well below one picosecond, with even the 506 kHz configuration reaching 500~fs compressed duration. At this point we emphasise that, throughout these tests, the cavity architecture remained the same in all respects other than fibre spool lengths.\looseness=-1


The experimental results above clearly illustrate that the new double-spool design can sustain much lower repetition rates that the single-spool configuration in \cite{ClaudeLargeChirpLaser}. Physically, we believe this can be explained simply in terms of the diminished optical nonlinearities ensuing from the specific distribution of fibre spools. In the single-spool design \cite{ClaudeLargeChirpLaser}, all the fibre is placed immediately after the gain fibre, where the intracavity intensity (hence, accumulation of nonlinearities) is large; in contrast, in the double-spool configuration, the second fibre spool is located right after the output coupler, where intracavity intensity is comparatively low. Interestingly, although the intensity would be even lower immediately after the band-pass filter [see Fig.~\ref{SchematicGCO}], we have not been able to mode-lock the laser when the second fibre spool is shifted to that position. A detailed study is beyond the scope of this Letter, but we speculate this may be because the pulse entering the main gain fibre would in that configuration be highly chirped, hindering its evolution. We reiterate that, because of limited availability of SMF, we have not been able to explore the ultimate repetition rate limits of the new double-spool design. However, we believe that our 506~kHz device is already nearing that limit, based on the deterioration of the spectral profile as well the increased difficulty in reaching the stable mode-locked regime.\looseness=-1

\begin{table}[htbp]
\centering
\caption{Design parameters and output characteristics for different laser configurations: $f_\mathrm{rep}$, repetition rate in kHz; $L_{1}$ and $L_{2}$, lengths of spools 1 and 2 in metres; $P_\mathrm{m}$ and $P_\mathrm{N}$, main and NALM pump powers in mW; $E_\mathrm{p}$, pulse energy in nJ; $P_\mathrm{av}$, average output power in mW; $\tau_\mathrm{uc}$, uncompressed duration in ps; $\tau_\mathrm{c}$, compressed duration in fs; $P_\mathrm{p}$, peak power in kW; for different repetition rates.}
\vspace{0.5em}
\begin{tabular}{cccccccccc}
\hline
$f_\mathrm{rep}$ & $L_{1}$ & $L_{2}$ & $P_\mathrm{m}$ & $P_\mathrm{N}$ & $E_\mathrm{p}$ & $P_\mathrm{av}$ & $\tau_\mathrm{uc}$ & $\tau_\mathrm{c}$ & $P_\mathrm{p}$\\
\hline
506 & 174 & 200 & 82 & 25 & 6.9 & 3.5 & 74 & 500 & 12 \\
672 & 114 & 160 & 71 & 25 & 7.1 & 4.8 & 88 & 346 & 21 \\
835 & 114 & 100 & 78 & 25 & 9.3 & 7.7 & 72 & 255 & 37 \\
908 & 100 & 100 & 71 & 30 & 9.0 & 8.2 & 70 & 221 & 41 \\
1128 & 60 & 100 & 88 & 25 & 7.4 & 8.3 & 57 & 216 & 34 \\

\hline
\end{tabular}
\label{TableSpecs}
\end{table}


To summarise, we have presented a new design for an enivironmentally stable Yb-doped mode-locked fibre laser that can reach ultra-low repetition rates. We have tested various configurations with different repetition rates, each of which has been able to produce nanojoule pulses that can be compressed to ultrashort durations. The performance characteristics of the realised configurations are summarised in Table~\ref{TableSpecs}. As can be seen, the lowest repetition rate that we have achieved is 506 kHz (limited by the available SMF), which to our knowledge is the lowest repetition rate ever reported for an environmentally-stable mode-locked fibre laser emitting broadband pulses. As the laser operates stably at these specifications, it can be considered a viable option as a seed laser in an amplified system for commercial micro-machining or medical imaging.\looseness=-1

\vspace{12pt}
\textbf{Funding.} Rutherford Discovery Fellowships of the Royal Society of New Zealand. Ministry of Business, Innovation and Employment, New Zealand.

\newcommand{\enquote}[1]{``#1''}

\end{document}